\documentclass[12pt]{iopart}

\def\be{\begin{equation}}
\def\en{\end{equation}}
\def\bea{\begin{eqnarray}}
\def\ena{\end{eqnarray}}
\def\bec{\begin{equation}\begin{array}{rcl}}
\def\p{\partial}

\def\gs{\gtrsim}
\def\ls{\lesssim}
\def\ve{\varepsilon}
\newcommand{\av}[1]{\langle{#1}\rangle}

\newcommand{\ten}[1]{\stackrel{\leftrightarrow}{\bi{#1}}}

\usepackage{graphicx}

\usepackage{iopams}  
\usepackage{cite}

\begin{document}

\title[Solvation Effects in Phase Transitions in Soft Matter]
{Solvation Effects in Phase Transitions in Soft Matter }

\author{Akira Onuki, Takeaki Araki and Ryuichi Okamoto}

\address
{Department of Physics, Kyoto University, Kyoto 606-8502,
Japan}
\ead{onuki@scphys.kyoto-u.ac.jp}

\begin{abstract}
Phase transitions in polar binary mixtures 
can be drastically altered  
even by  a small amount of salt. 
This is because  
the preferential solvation 
 strongly depends   on the ambient composition. 
Together with a summary 
of our research in this problem,  
 we present some detailed results 
 on the role of 
 antagonistic salt composed of hydrophilic and hydrophobic ions. 
These ions  tend to segregate at liquid-liquid interfaces  
and  selectively couple to water-rich and oil-rich composition 
fluctuations, leading to mesophase formation. 
In our two-dimensional simulation, 
the corasening of the domain structures 
can be stopped or slowed down,  
depending on the interaction parameter (or the 
temperature) and the salt density. 
We realize stripe patterns at the critical 
composition and droplet patterns at  
off-critical compositions. 
In the latter case,  
charged   droplets emerge 
with considerable size dispersity  
in a percolated  region. 
We also give the structure factors among 
the ions,  accounting for the Coulomb interaction and 
the  solvation interaction mediated by 
the composition fluctuations.  
\\
(Some figures in this article are in colour only in the electronic version)

\end{abstract}

\maketitle

\section{Introduction}
\setcounter{equation}{0}

In many soft materials, 
phase transitions and structure formations 
occur in the presence of charged objects  including 
small ions,  charged  colloids, and polyelectrolytes   
 in  polar fluids (mostly in  water)  \cite{Levin,Holm,Rubinstein}. 
Much attention has been paid 
to the consequences of the long-range  
 Coulomb interaction,  but not enough attention 
 has yet been paid to 
the solvation interaction among 
charged objects and polar molecules \cite{Is,Marcus,Ohtaki,Gut}. 
In the theoretical literature in  physics, 
the solvation interaction has been mostly  neglected 
in analytic theories   and  numerical simulations. 
In real aqueous systems, a number of molecular 
interactions 
are operative  at short 
distances under the influence of the  long-range  electric 
potential \cite{Is}. In particular, the 
  hydrogen bonding is of primary importance  in the 
phase behavior in  aqueous systems, 
as studied   in neutral polymer solutions 
\cite{TanakaF}.  However,   small ions like Na$^+$ 
form a solvation shell 
composed of  several water molecules 
and are  hydrophilic, around which     
the energy of the ion-dipole interaction typically   
exceeds    the energy of the surrounding 
hydrogen bonds.  On the contrary, some 
 neutral or ionized particles 
dislike to be in contact 
with water molecules and become hydrophobic. 
Hydrophilic and hydrophobic 
particles deform the surrounding hydrogen bonding  
in different manners \cite{Is}. 
Thus the solvation is highly 
complex in aqueous mixtures.  
In this paper, we will show 
that   the preferential (or selective) solvation of ions 
 in a mixture solvent 
  can  drastically 
influence   the phase transition  behavior.

Nabutovskii {\it et al.}\cite{Russia} 
 found a possibility of 
mesophases  in electrolytes 
in the presence of  a coupling between the composition and 
the  charge density in the free energy. 
In our theory  of electrolyte \cite{OnukiJCP04,OnukiPRE}, 
such a coupling originates 
from  the solvation  in polar  fluids and 
its magnitude is  exceedingly high   
in many real systems.
Recently, including the  solvation effect 
in mixture solvents, several theoretical 
groups  have  examined the ion effects  in electrolytes 
\cite{OnukiJCP04,OnukiPRE,OnukiJCP,Tsori,Roij,Andelman1,Araki,Mo,An,Daan}, 
polyelctrolytes\cite{Onuki-Okamoto,Oka}, 
and ionic surfactants \cite{OnukiEPL} 
using  phenomenological models. 
A review on the static properties in this line 
was presented in Ref.$[23]$. In the dynamics, a 
number of  problems 
still remain not well studied  
 \cite{Araki,Daan}. On the other hand, 
a large number of microscopic simulations 
have been performed, for example,  to calculate 
the ion distributions  near 
water-air interfaces \cite{Trobias} 
and  solid surfaces \cite{Netz,Ju}.

In this paper, 
we  are  particularly interested in statics 
and dynamics of  phase transitions 
in aqueous mixtures   with antagonistic salt, which 
is  composed of 
hydrophilic and hydrophobic ions. 
While the surface tension of liquid-liquid interfaces 
is  known to be   slightly increased by    
 hydrophilic ion pairs \cite{Onsager,Levin-Flores}, 
 it can be dramatically decreased by    
antagonistic salt as observed \cite{Reid,Luo}. 
Hydrophilic and hydrophobic ions segregate  
at liquid-liquid  interfaces on the scale of the Debye 
 screening length $\kappa^{-1}$, 
resulting in   a large electric 
double layer \cite{OnukiPRE,OnukiJCP,Araki}. 
They also interact differently with water-rich and oil-rich 
composition fluctuations. 
Mesophases (charge density waves)  then appear   
 near   the solvent 
criticality for sufficiently large  solvation 
asymmetry between the cations and the anions.   

We mention   recent  
experiments where antagonistic salt was used. 
(i) Luo {\it et al} observed a unique  
interfacial ion distribution   
by x-ray   reflectivity experiment \cite{Luo}.   
(ii) Sadakae {\it et al.} 
\cite{Sadakane} added a small amount of 
sodium tetraphenylborate 
NaBPh$_{4}$     to a  near-critical 
mixture of D$_2$O 
and tri-methylpyridine (3MP). This salt dissociates into  
 hydrophilic  Na$^+$ 
and hydrophobic  BPh$_{4}^-$. The latter anion     
consists  of  four phenyl rings bonded to an ionized 
boron.  They found  a  peak   
at an intermediate wave number 
$q_m$($\sim 0.1~$\AA$^{-1}\sim \kappa$) 
in the intensity  of small-angle neutron scattering. 
The peak height  was much 
enhanced with formation of periodic structures.    
Moreover, they observed  multi-lamellar (onion) 
structures   at  small volume fractions of 3MP 
(in D$_2$O-rich solvent)  far from 
the critical point \cite{SadakanePRL}, 
where BPh$_{4}^-$  and solvating 3MP form charged lamellae. 
These  findings  demonstrate 
  very strong hydrophobicity of BPh$_{4}^-$, 
  which was  
  effective close to the solvent criticality \cite{Sadakane} 
  and far from it  \cite{SadakanePRL}. 
(iii) Another intersting phenomenon 
is    spontaneous emulsification 
(formation of small water droplets) 
at a water-nitrobenzene(NB) interface \cite{Aoki,Poland}.
It was observed when a large water droplet was pushed  into 
a cell  containing NB and  antagonistic salt 
(tetraalkylammonium chloride). This 
 instability was caused by 
ion transport through the interface.

The organization of this paper is as follows. 
In Section 2,   we will present   a  summary 
of the solvation effects, particulaly focusing on 
 the preferential solvation.   
In Section 3,  the Ginzburg-Landau model 
and dynamic equations for such systems will be given.  
As a new result, 
the surface tension of  binary mixtures 
containing an antagonistic salt will  be shown to 
decrease to zero with increasing its   density.  
In Section 4, 
we  will  numerically examine 
the mesophase formation with an antagonistic salt 
in two dimensions (2D).

\section{Solvation effects}
This section is an introduction 
of the preferential solvation. It 
provides a starting point 
of our theory in Section 3.

\subsection{ Solvation shell and Born theory} 

In polar liquids,  a number of 
poalr molecules are attached to 
 a hydrophilic ion  such as Na$^+$, Li$^+$, 
 or Ca$^{2+}$ to 
 form  a solvation shell due to the ion-dipole 
 interaction \cite{Is,Wakisaka} (see figure 1).  
On the other hand,   
hydrophobic particles 
interact with water molecules repulsively, 
so they  tend to form 
aggregates in water and are 
more soluble in oil than in water.  
We  may  consider  
mixture solvents  composed of polar and less polar 
components, where hydrophilic (hydrophobic) ions 
are surrounded by molecules of 
the more (less) polar component  \cite{Marcus,Gut} 
(see the right panel of figure 1). 
That is, solvation occurs preferentially or selectively in 
mixture solvents.   There can be a variety of 
mixture solvents and, in this paper,  
we call the two components simply as 
water and oil.  Note that 
solvation is also called 
hydration for water (or heavy water) 
and  for aqueous mixtures containing water.

In this paper, we introduce a solvation chemical potential 
$\mu_{\rm sol}^i(\phi)$ for   each  hydrophilic or hydrophobic 
charged  particle, where $i$ represents the particle 
species and $\phi$ is the water composition. 
It is a thermodynamic 
quantity obtained  after 
the statistical average over the thermal fluctuations. 
With formation of a well-defined 
solvation shell, the typical 
magnitude of $\mu_{\rm sol}^i$ 
much exceeds the thermal energy $k_BT$. 
In his  original theory,  Born    took  into account 
the polarization of polar molecules 
around a spherical  ion using a continuum 
theory of electrostatics  
 \cite{Born}. 
It is the space integral of the electrostatic energy 
$\ve {\bi E}^2/8\pi$, 
where ${\bi E}=-\nabla \Phi$  in terms of 
the potential $\Phi$. Here $\Phi=Z_i e/\ve r$ 
depends  on the distance $r$ from 
the ion center  with $Z_ie$ being the  charge. 
The classic  Born formula thus obtained is written as  
\be  
(\mu_{\rm sol}^i)_{\rm Born} 
=  {Z_i^2e^2}/{2R_{\rm ion}^i\ve} 
=  k_{B}T 
Z_i^2\ell_B/{2R_{\rm ion}^i },
\en 
which is applicable for hydrophilic small ions. 
Here the linear dielectric constant $\ve$ 
is used, though the electric field in the vicinity of 
a small  ion is strong beyond 
the linear response level (leading to dielectric saturation). 
The  space integral is in the region 
$R_{\rm ion}^i<r<\infty$, where the lower cut-off  
  $R_{\rm ion}^i$ is called the Born radius of order 
  $1{\rm \AA}$ for small metallic ions \cite{Is,RMarcus}.  
   In terms of the 
Bjerrum length $\ell_{{\rm B}}= e^2/\ve k_{B}T$, 
  the Born form $(\mu_{\rm sol}^i)_{\rm Born}$ 
  itself  considerably exceeds 
   the thermal energy $k_BT$ 
 for $R_{\rm ion}^i<\ell_{{\rm B}}$. 
 Here $\ell_{\rm B}$ is $7 {\rm \AA}$ 
 for room-temperature  water and 
 is longer for less polar solvents.  
The Born formula  is in accord with 
the fact that the solvation is stronger for smaller ions. 
It also indicates that the solvation is much intensified 
for multivalent ions such as Ca$^{2+}$ or Al$^{3+}$. 
For mixture solvents,  
the dielectric constant 
$\ve=\ve(\phi)$ depends on the 
composition $\phi$ of the more polar 
component A \cite{Debye}, changing 
 from   $\ve_{\rm B}$ of the less polar component B to  
 $\ve_{\rm A}$ of the more polar component A. Since the Born 
expression is inversely proportional to $\ve(\phi)$,  
 it indicates   strong preferential solvation.  
 However, the Born formula is 
 known to be very  crude neglecting 
 the formation of the shell structure, the density and 
 composition changes (electrostriction), and 
 the nonlinear dielectric effect. 
Moreover,  the solvation should  
be strongly coupled with  the surrounding 
hydrogen bonds  in aqueous mixtures.

\begin{figure}
\begin{center}
 \includegraphics[scale=0.32, bb= 0 0 760 432]{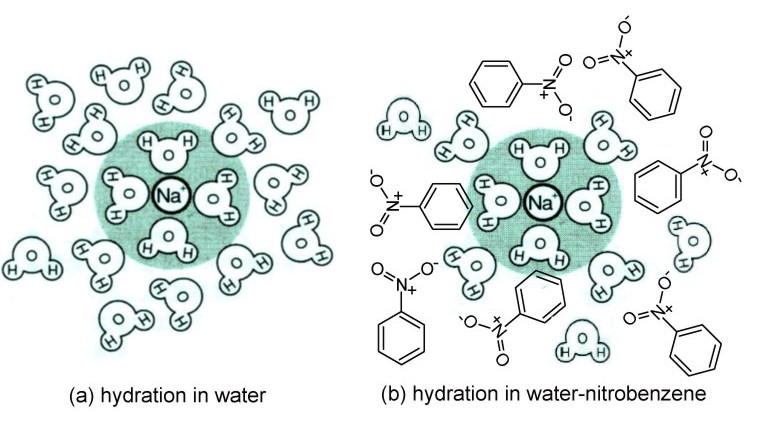} 
\end{center}
 \caption{Illustration of 
hydration of Na$^+$  surrounded 
by a shell composed of water molecules in 
(a) pure water and (b) water-nitrobenzene.  
 }
\end{figure}

We give  a simple 
argument on the condition of the formation of 
a well-defined solvation shell composed of 
several polar molecules. 
Let the solvent molecular size 
$a= v_0^{1/3}$ be longer than the ion radius 
 $\sim R_{\rm ion}^i$ (where $a \sim 3{\rm \AA}$ 
 for water).    For   single component polar liquids,  
 we compare the electrostatic energy 
in the region $r\sim a$ 
and the thermal energy  $k_BT$ 
to find the criterion,  
\be 
A_0\equiv  Z^2e^2/8\pi \ve k_BT a \gs  1.
\en 
For mixture solvents, a phenomenological 
theory yields  the criterion \cite{OnukiJCP04}, 
\be 
A \equiv  Z^2e^2 \ve_1/8\pi \ve^2 k_BT a \gs  1,
\en 
for the formation of 
a well-defined  preferential 
shell.  Here 
 the  molecular volumes of the two componenta are 
assumed to be 
 of the same order $v_0$ and we define 
\be 
\ve_1= \p \ve(\phi)/\p\phi,
\en 
where the derivative is at fixed $T$ and $p$.
  In these relations, we have estimated 
 the binding energy $\epsilon_b$ of a polar molecule 
 to an ion  to be 
 $Ak_BT$ or $A_0 k_BT$. 
Since $\epsilon_b$ is a microscopic energy, 
these estimations should be very crude (as well as the 
Born formula). In this theory, the 
 strong solvation regime is realized 
for $A_0\gg 1$ or for $A\gg 1$.  If the influence 
on the hydrogen bonding network is neglected,  the 
radius $R_{\rm shell}^i$ of a 
solvation  shell   grows  
as $A_0^{1/4}a$ or as $A^{1/4}a$ 
from the decay of the electrostatic energy density 
 $\propto   r^{-4}$.  Here it is worth noting that 
  the shell radius 
$R_{\rm shell}^i$ plays the role of 
the hydrodynamic radius  $R_h^i$ in 
the  diffusion  constant $D_i
= k_BT/6\pi \eta R_h^i$ with $\eta$  being the shear viscosity.  
As a result, ions with smaller molecular radii ($\sim R_{\rm ion}^i$) 
 have larger  hydrodynamic radii, 
 which is explained from 
 the relation  $R_h^i \cong R_{\rm shell}^i$ \cite{Ham}.  
 For example, Li$^+$ ions are smaller than Na$^+$ ions, 
 but the diffusion constant of 
Li$^+$ is smaller than that of Na$^+$ in water.

In electrochemistry, much attention 
has been paid  to the ion distribution 
 and the electric potential difference 
 across a liquid-liquid  interface \cite{Ham,Hung}. 
Remarkably, water molecules solvating 
a hydrophilic ion in the water-rich 
region remain attached to 
the ion even when 
the ion crosses an interface and 
moves  into the oil-rich region. 
 Osakai {\it et al}   \cite{Osakai}  
measured the amount of water molecules 
  extracted together with hydrophilic ions  in 
a nitrobenzene(NB)-rich phase with a 
small water composition ($\sim 0.168$M). 
In NB-water  at room temperatures,  
the number of  coextracted water 
molecules in a NB-rich phase 
was estimated to be  4 for Na$^+$, 
6 for Li$^+$, and 15 for Ca$^{2+}$. 
Furthermore, using proton NMR spectroscopy,   
Osakai {\it et al.} \cite{proton} 
studied successive formation 
of complex structures of anions (such as Cl$^-$ and Br$^-$) 
and water molecules 
by gradually increasing the water composition in NB.

These experiments  demonstrte  
that the  solvation shell 
 composed of water molecules is  stable even when 
the ambient water composition 
is much decreased. This is a remarkable 
result  suggesting  a number of important 
consequences. For 
 example, with  a small amount of 
 hydrophilic  ions in oil, let us  
 gradually increase the  water composition  
 $\phi$ from zero.  If the binding energy $\epsilon_b$ 
 of a water molecule to such an ion 
 is much larger than $k_BT$, we may argue 
  that  a hydration shell is formed for 
\cite{Oka}
\be 
\phi> \phi_{\rm sol} \sim  \exp(-\epsilon_b/k_BT),
\en 
where the crossover composition 
$ \phi_{\rm sol}$ is very small for strongly 
hydrophilic ions. 
The solvation chemical 
potential $\mu_{\rm sol}^i(\phi)$ 
of   hydrophilic ions  in water-oil 
should decrease considerably  in the narrow dilute 
range $0<\phi< \phi_{\rm sol}$.
Its  variation  should be milder in the range (2.5). 
The solbility of such ions 
should also 
increase abruptly in the same  narrow range of $\phi$ 
with addition of water to oil. Therefore, 
solubility measurements of hydrophilic ions are informative 
at small water compositions.

\subsection{Gibbs transfer free energy and 
 solvation chemical potentials }

We consider  a liquid-liquid  interface  between  
a  polar phase $\alpha$ and a less polar 
phase $\beta$ with bulk compositions $\phi_\alpha$ 
and $\phi_\beta$,  across which there arises a 
difference in  $\mu_{\rm sol}^i$ due to 
its composition dependence:  
\be 
\Delta\mu_{\alpha\beta}^{i}
= \mu_{\rm sol}^{i\alpha}-\mu_{\rm sol}^{i\beta}, 
\en 
where $\mu_{\rm sol}^{iK}$ ($K=\alpha,\beta$) are
the bulk values of the solvent chemical 
potential of  species $i$ in the two phases. 
In electrochemistry \cite{Hung,Osakai}, 
the difference of the solvation free energies  
$\Delta G_{\alpha\beta}^{i}$ between two phases  
  have been called the standard 
Gibbs transfer energy. (Since $\Delta G_{\alpha\beta}^{i}$ 
is usually measured in units of 
kJ per mole, dividing it   by the Avogadro number gives   
 $\Delta\mu_{\alpha\beta}^{i}$ per ion). 
  It is well-known that 
if there are differences 
among $\Delta\mu_{\alpha\beta}^{i}$ ($i=1,2,\cdots)$, 
 an electric double layer emerges 
 at the interface, giving rise to an electric  potential jump 
 $
 \Delta\Phi=\Phi_\alpha-\Phi_\beta,
$    
across the interface in equilibrium (called the Galvani potential difference).
Similar potential differences also appear 
at liquid-solid interfaces (electrodes) \cite{Ham}. 
It is also worth noting that 
there can   be an 
electric potential difference between two 
liquids due to orientation of molecular dipoles 
even without added ions, for which  
see  Ref.$[43]$  for water-hexane.

To examine the above effect, let us 
suppose only two species of ions ($i=1, 2$) 
with charges $Z_1e$ and $Z_2e$ ($Z_1>0$ and $Z_2<0$). 
At sufficiently low ion densities, 
the  ion chemical potentials are expressed as 
\cite{Ham,Hung} 
\be 
\mu_i= k_BT \ln (n_i\lambda_i^3) 
+ Z_ie\Phi+ \mu_{\rm sol}^i (\phi),
\en 
for $i=1,2$. 
The $\lambda_i$ is the thermal de Broglie length 
(but is an irrelevant constant in the following). From 
the charge neutrality in the bulk regions, 
the bulk ion densities $n_{1\alpha}$, $n_{1\beta}$, $n_{2\alpha}$, 
and $n_{2\beta}$ satisfy
\be 
Z_1n_{1\alpha}+Z_2n_{2\alpha}=0, \quad 
   Z_1n_{1\beta}+Z_2n_{2\beta}=0.
\en    
The continuity of $\mu_i$ across the interface yields 
\be 
k_BT \ln (n_{i\alpha}/n_{i\beta})+ Z_i e\Delta\Phi 
+ \Delta\mu_{\alpha\beta}^{i}
=0, 
\en 
for $i=1,2$. Then the  Galvani potential difference 
 is expressed as \cite{Hung,OnukiPRE}
\be
\Delta \Phi=\frac{\Delta\mu_{\alpha\beta}^2 
-\Delta\mu_{\alpha\beta}^{1}}{e(Z_1+|Z_2|)},
\en 
The ion densities in the bulk two phases 
are related by  
\be 
\ln \bigg(\frac{n_{1\alpha}}{n_{1\beta}}\bigg )
= \ln \bigg ( \frac{n_{2\alpha}}{n_{2\beta}}\bigg ) = 
\frac{|Z_2| {\Delta\mu_{\alpha\beta}^{1}} +
Z_1 \Delta\mu_{\alpha\beta}^{2} }{(Z_1+|Z_2|)k_BT} ,
\en

Let us take the $z$ axis perpendicular to a liquid-liquid 
interface. 
While the interface thickness is 
given by the correlation length $\xi$, 
the potential  
$\Phi(z)$ changes on the scale of  
the Debye-H$\ddot{\rm u}$ckel 
screening length,  $\kappa_\alpha^{-1}$ in phase  $\alpha$  
and $\kappa_\beta^{-1}$ in phase  $\beta$. 
As a result,  $\Phi$ changes from 
$\Phi_\alpha$ to $\Phi_\beta$ on  the spatial scale  
 of $\kappa_\alpha^{-1}+\kappa_\beta^{-1}$, which becomes 
very long as the ion densities determined by equation (2.11) 
become very small in one of the two phases 
 \cite{OnukiPRE}.   Remarkably,  
$\Delta\Phi$ in equation (2.10) is  independent of the ion densities. 
It is typically of order 
$10k_{B}T/e (=258$mV)  for 
$\Delta\mu_{\alpha\beta}^2\neq 
\Delta\mu_{\alpha\beta}^{1}$. It  vanishes 
for symmetric ion pairs  with $\Delta\mu_{\alpha\beta}^2 
=\Delta\mu_{\alpha\beta}^{1}$. As equation (2.11) indicates, 
the ion densities in the two phases 
are very different in most  cases. 
For example, if 
  $\Delta\mu_{\alpha\beta}^1/k_{B}T= 
\Delta\mu_{\alpha\beta}^2/k_{B}T=10$   in the monovalent case, 
the common ratio  ${n_{1\beta}}/{n_{1\alpha}}
={n_{2\beta}}/{n_{2\alpha}}$  
becomes  $e^{-10}=4.5\times 10^{-5}$. 
If $\kappa_\beta^{-1}$ 
is macroscopic (as in air in contact with 
salted water),  $\Delta\Phi$ may not be 
observable.

In  the literature,  data of  
$\Delta G_{\alpha\beta}^i$ 
on water-nitrobenzene at room temperatures are available 
\cite{Hung,Osakai}, 
where water and NB are strongly segregated but 
a considerable amount water is present in the NB-rich 
phase ($0.168$M).  
For aqueous mixtures 
with $\alpha$ being the water-rich phase, 
$\Delta G_{\alpha\beta}^i$ is negative 
for hydrophilic ions and positive  for hydrophobic  ions. 
For water-nitrobenzene  \cite{Hung}, 
we have   $\Delta\mu_{\alpha\beta}^i/k_{B}T= $  
-13.6 for Na$^+$, -15.3 for Li$^+$,  -26.9  for Ca$^{2+}$, 
-11.3 for Br$^-$, and -7.46 for I$^-$ as examples of 
hydrophilic ions, while we have 
$\Delta\mu_{\alpha\beta}^i/k_{B}T= 14.4$ for  hydrophobic  
BPh$_4^-$.   
In these experiments,  a hydration shell 
should have been formed around each hydrophilic ion 
even in the water-poor phase $\beta$.  
Equation (2.5) then suggests 
$\phi_\beta>\phi_{\rm sol}$. Though more data 
are needed, those  of $\Delta G^i_{\alpha\beta}$ 
for NB-water indicates  the 
composition-dependence of  $\mu^i_{\rm sol}(\phi)$ 
 quantitatively, which is very strong even after the 
formation of solvation shells.

\setcounter{equation}{0}
\section{Ginzburg-Landau Theory}
We will first give  a short summary of our previous results 
in the Ginzburg-Landau scheme in the presence of 
the preferential  solvation in Subsections 3.1-3.3. 
New results will then follow  
on the structure factors among ions,  
the surface tension behavior with antagonistic salt, 
 and the phase diagram. 
We treat  two species of monovalent ions. 
However, for three species of ions, 
 we obtain more complex 
 ionic  distributions near 
liquid-liquid  interfaces \cite{Luo,OnukiJCP}   
and solid surfaces \cite{Ju}.

\subsection{Electrostatic and solvation interactions}

We consider a polar binary 
 mixture  (water-oil)  
containing a small amount of monovalent 
 salt ($Z_1=1$ and $Z_2=-1$). 
The ions  are sufficiently dilute and 
 their  volume fraction is negligible.  
Hereafter the Boltzmann constant will be set equal to unity. 
Neglecting the image interaction, 
we  assume the  free energy density 
\cite{OnukiJCP04,OnukiPRE},  
\begin{eqnarray}
&&{f} = f_0(\phi,T) +
\frac{T}{2}C|\nabla\phi|^2+ \frac{\varepsilon}{8\pi }{\bi E}^2
\nonumber\\
&& + T\sum_i
\bigg [\ln (n_i\lambda_i^3) -1-  g_i \phi\bigg]n_i.  
\end{eqnarray} 
The first term 
$f_0$ is the chemical part. 
In our simulation we use the Bragg-Williams form,    
\be 
\frac{v_0}{T} f_0  =   
 \phi \ln\phi + (1-\phi)\ln (1-\phi) 
+ \chi \phi (1-\phi),   
\en 
where $v_0$ is the solvent molecular volume 
common to the two components and  $\chi=\chi(T)$ 
depends   on  $T$ (at constant pressure). 
Here $\chi=2$ at the 
 solvent critical temperature $T=T_c$ without ions 
 and $\chi-2 \propto T-T_c$ for small $T-T_c$. 
The second term is  the gradient part with 
$C$ being a constant. The  third  term is the 
electrostatic free energy, where 
${\bi E}=-\nabla\Phi$ is the electric field and the potential $\Phi$  
is produced by the charge density 
$\rho= e(n_1-n_2)$ via  the Poisson equation,  
\be 
-\nabla\cdot\ve(\phi)\nabla \Phi=  4\pi \rho.
\en 
The dielectric constant 
  $\ve(\phi)$ can   depend  on 
the composition $\phi$ \cite{Debye}. In our previous work 
the linear composition dependence  
\be 
\ve(\phi)=\ve_0 + \ve_1 \phi
\en 
has been assumed, where $\ve_0$ and 
 $\ve_1$ are constants. 
The last term in the right hand side of 
equation (3.1) consists of the entropic part 
and  the solvation part 
of the ions, with  $\lambda_i$ being the thermal 
de Broglie length. The  solvation terms $(\propto g_i)$  
follow if   $\mu_{\rm sol}^i(\phi)$ 
 depend on $\phi$ linearly as  
\be 
\mu_{\rm sol}^i(\phi) ={\rm const.}  -Tg_i\phi, 
\en  
where  the constant term gives 
rise to a  contribution linear in $n_i$ in $f$  
and is irrelevant, while the second term $(\propto g_i$) 
 yields the solvation  coupling  in $f$. 
In this approximation, the parameter $g_i$ 
represents the strength of the preferential 
solvation of ion species $i$. 
The difference of the solvation 
chemical potential in two-phase coexistence  in equation (2.6) 
 is given by  
\be 
\Delta\mu_{\rm sol}^i= 
Tg_i\Delta\phi,
\en 
where $\Delta\phi$ is the 
composition difference. Thus  $g_i>0$ 
for hydrophilic ions and $g_i<0$ 
for hydrophobic ions. The  discussion in 
the subsection 2.2 indicates   
 $g_i \sim 15$  
 for Na$^+$ ions and 
 $g_i \sim -15$  
for  BPh$_4^-$ in water-nitrobenzene at 300K. 

The linear form equation (3.5) 
is adopted for the mathematical 
simplicity and should not be taken too seriously. 
Moreover, around equation (2.5), we predict a steep 
drop of $\mu_{\rm sol}^i(\phi)$ 
in the range $0<\phi<\phi_{\rm sol}$ 
for hydrophilic ions with formation of a 
hydration  shell. Thus equation (3.5) 
is applicable for $\phi>\phi_{\rm sol}$.  
In  addition, 
in our previous papers \cite{OnukiPRE,OnukiJCP}, 
the image interaction was also included in the free energy, 
but it is neglected in this paper for simplicity.

For  a monovalent antagonistic salt,  
figure 2 gives  
equilibrium interface   profiles 
with $g_1=-g_2=10$ at  $\chi=3$  
(see Ref.$[11]$ for more details), where 
we can see a marked electric double layer. 
The bulk cation densities in the two sides,  
  $n_{1\alpha}$ and $ n_{1\beta}$,  coincide 
 from equation (2.8) and is set equal to 
$2\times 10^{-4}v_0^{-1}$.  
The interface thickness 
$\xi$ is of order $5a$, 
while $\Phi(z)$ changes on the scale of 
$\kappa_\alpha^{-1}\sim \kappa_\beta^{-1}\sim 10a$. 
The surface tension  change   due to salt is 
$\Delta\sigma=\sigma-\sigma_0= 
 -0.041Ta^{-2}$  (where  $\sigma_0= 0.497 T/a^2$ 
 without salt) and 
the surface adsorption of ions is  
$\Gamma= 0.014a^{-2}$ (see equation (3.33)).  
If   $g_1=-g_2$, the  
electric double layer is increasingly 
enhanced with further 
increasing $g_1$ and/or $n_{1\alpha}$, 
eventually leading to vanishing of 
the surface tension (see figure 3).  
We can understand these features 
by solving the nonlinear Poisson-Boltzmann equation 
 \cite{OnukiJCP}.
See also a similar recent calculation 
of the interface profiles \cite{Daan}. 
The ionic  interfacial 
 distribution was measured  for a ternary, 
 antagonistic salt in water-NB  
in the experiment by Luo {\it et al.}\cite{Luo} 
and was theoretically explained in our 
previous paper \cite{OnukiJCP}.

\begin{figure}
\begin{center}
\includegraphics[scale=0.45, bb=0 0 408 643]{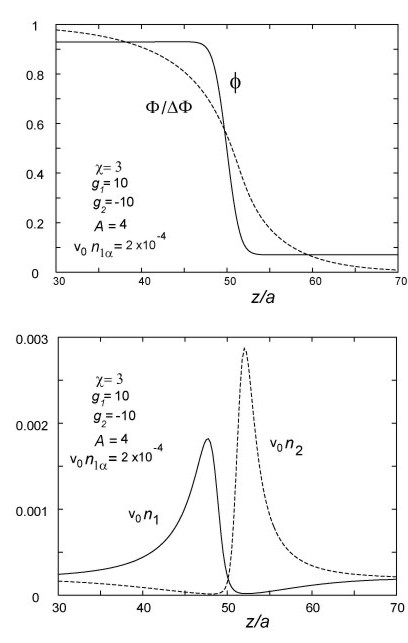}
\end{center}
\caption{ 
Normalized potential 
$\Phi(z)/\Delta\Phi$  and composition $\phi(z)$ 
(upper panel), and normalized ion densities 
$v_0 n_1(z)$  and $v_0n_2(z)$ (lower panel) 
(taken from Ref.$[11]$),
where $\chi =3, g_1=g_2=10$, 
and $v_0n_{1\alpha} =v_0n_{1\beta}
=2\times 10^{-4}$. Here $A=4$ means $e^2/a\ve_0T= 8.5$.  
Hydrophilic cations (hydrophobic anions) tend to be 
on the left (right) near the interface, 
resulting in a large electric double layer.  
The asymmetry of the two curves of $n_1$ and $n_2$ is due to 
the $\phi$-dependence of $\ve(\phi)$ ($\ve_1/\ve_0=4/3$ 
here). 
}
\end{figure}

\subsection{Dynamic equations}

We present the dynamic equations 
for $\phi$, $n_1$, $n_2$, 
and the velocity field $\bi v$ \cite{Araki,Onukibook}.
The fluid is incompressible and isothermal, so 
 we assume  
$\nabla\cdot{\bi v}=0$  
and treat 
the mass density $\rho_0$  
and the temperature $T$ as constants. 
Then  $\phi$ and $n_i$ obey  
\begin{eqnarray}
&&\frac{\partial\phi}{\partial t}+
\nabla\cdot(\phi\mbox{\boldmath $v$})= 
\frac{L_0}{T} \nabla^2\frac{\delta F}{\delta \phi},\\
&&\frac{\partial n_i}{\partial t}+
\nabla\cdot(n_i\mbox{\boldmath $v$})= \frac{D_i}{T}
\nabla\cdot n_i\nabla \frac{\delta F}{\delta n_i} \nonumber\\
&&=D_i\nabla\cdot\bigg[{\nabla{n_i}}
\mp \frac{e}{T}  n_i{\bi E} -g_in_i\nabla\phi\bigg],
\end{eqnarray}
 The kinetic coefficient  $L_0$  
   and  the ion diffusion constants 
    $D_1$ and $D_2$ are constants. 
The symbol $\pm$ in the second line of equation (3.8) 
denotes $-$ for the cations $(i=1)$ and 
$+$ for the anions ($i=2$).
The momentum equation is expressed as 
\be 
\rho_0  \frac{\partial{\bi v}}{\partial t}=
-\nabla p_1 + \eta\nabla^2{\bi v} - \nabla\cdot 
 {\ten \Pi}  ,
\en
where  $p_1$  ensures $\nabla\cdot{\bi v}=0$, $\eta$ is the 
shear viscosity, and 
${\ten \Pi}=\{\Pi_{\alpha\beta}\}$ ($\alpha,\beta=x,y,z$) 
is the reversible stress tensor of  the form, 
\be 
\Pi_{\alpha\beta}=
{T}C  \nabla_\alpha\phi \nabla_\beta\phi
- \frac{\ve}{4\pi}E_\alpha E_\beta. 
\en 
with $\nabla_\alpha=\p/\p x_\alpha$. 
It follows the relation  
$ 
\nabla \cdot {\ten{\Pi}} = 
 \phi \nabla (\delta F/\Delta \phi) 
  +\sum_i  n_i \nabla (\delta F/\delta n_i),  
$  which is needed for the self-consistency of 
our model dynamic equations \cite{Araki,Onukibook}

In numerically integrating  the 
dynamic equations, we may use the following 
two approximations \cite{Araki}. First, in the 
 Stokes approximation,  we set  the 
 right hand side of equation (3.9) equal to zero. 
 Then $\bi v$ is expressed in terms of the 
 Oseen tensor inversely proportional to the viscosity $\eta$ 
 \cite{Onukibook}. 
Second,  in the quasi-static approximation, 
 we assume that the 
characteristic ion diffusion time 
is much faster than the the relaxation time of 
the composition. 
This is justified  near the 
solvent critical point and in describing slow 
domain growth. 
In such cases, the ion distributions are 
expressed in terms of  $\phi$ and $\Phi$ 
in the modified Poisson-Boltzmann relation \cite{OnukiPRE},  
\be 
n_i=  n_i^0 \exp[g_i \phi \mp e\Phi/T].
\en
This form was also used in analysis of 
experimental data  \cite{Luo} 
and in that of molecular dynamics 
simulation \cite{Netz}.
The coefficients 
 $n_i^0$ are  determined 
from the conservation of the ion numbers, 
\be  
\av{n_i}= 
\int d{\bi r}n_i({\bi r})=n_0,
\en  
where $\av{\cdots}=V^{-1}\int d{\bi r}(\cdots)$ 
denotes the space average 
with $V$ being the cell volume. The average density  $n_0
=\av{n_1}=\av{n_2}$ 
is a given constant.  It  will be measured
 in units of $v_0^{-1}$ or the normalized 
 density $v_0n_0$ will be used with $v_0$ 
 appearing in equation (3.2).

\subsection{Structure factors  and effective 
ion-ion interaction  in one-phase states}

We suppose small deviations 
$\delta\phi({\bi r})= \phi({\bi r})-\av{\phi}$ 
and $\delta n_i({\bi r})= n_i({\bi r})-n_0$ 
of the composition  and the ion densities 
in one-phase states. Their equilibrium distributions 
are treated to be Gaussian   in the mean field theory. 
Then, from the free energy (equation (3.1)),  
we may readily calculate the following 
 structure factors, 
\bea 
S(q)&=& {\av{|\phi_{\small{\bi q}}|^2}}_{\rm e},\quad 
G_{ij}(q)= {\av{n_{\small{i\bi q}} n_{j\small{\bi q}}^*   }}_{\rm e}/ n_0^, 
\nonumber\\ 
C(q)&=&
{\av{|\rho_{\small{\bi q}}|^2}}_{\rm e}/e^2n_0,
\ena 
where  $\phi_{\small{\bi q}}$, 
 $n_{i\small{\bi q}}$, and $\rho_{\small{\bi q}}$  
 are the Fourier components  of the composition 
 $\phi$,  the ion densities $n_i$, 
 and the charge density $\rho=e(n_1-n_2)$ 
with wave vector $\bi q$ and 
 $\av{\cdots}_{\rm e}$ 
denotes the thermal average. It  follows 
the relation, 
\be 
C(q)=G_{11}(q)+G_{22}(q)-2G_{12}(q).
\en    
The $\phi$-dependence 
of the dielectric constant $\ve$ is negligible for small 
fluctuations. We introduce the  Bjerrum length $\ell_B$ 
and the Debye wave number $\kappa$  by  
\be 
 \ell_B=  e^2/\ve T, \quad 
\kappa= (8\pi  \ell_Bn_0)^{1/2}.
\en

After elimination the ion fluctuations, 
the second order contribution of the free energy 
is written as $\delta F^{(2)}= T \sum_{\bi q}
|\phi_{\small{\bi q}}|^2/2VS(q)$. 
The inverse of $S(q)$ is written as 
\cite{OnukiJCP04,OnukiPRE} 
\be
\frac{1}{S(q)}
={\bar r}- (g_1+g_2)^2\frac{n_0}{2}  + Cq^2 
\bigg[1-  \frac{\gamma_{\rm p}^2\kappa^2}{q^2+\kappa^2}\bigg] , 
\en
where $\bar r$ is related to the second derivative of 
$f_0$ and use of equation (3.2) yields 
\be 
{\bar r}=f_0''/T =  
 v_0^{-1}[1/{\phi(1-\phi)}-2\chi]. 
\en  
The parameter $\gamma_{\rm p}$ 
represents asymmetry of the solvation 
of the two ion species and is defined by 
\be
\gamma_{\rm p}= (16 \pi C \ell_{B})^{-1/2} 
|g_1-g_2|
\en 
Remarkably, $\gamma_{\rm p}$ is independent of the salt density. 
Note that the structure factor for  weakly ionized 
polyelectrolytes has the same 
form \cite{Holm,Rubinstein}, where 
the solvation has been neglected, however. 
Recently, we have presented a generalized form of 
the structure factor 
for polyelectrolytes 
 including the solvation interaction and the ionization 
 fluctuations \cite{Onuki-Okamoto}.

We explain implications of 
the structure factor (equation (3.16)).  
(i) The second term   gives rise to 
a shift of the spinodal curve.  
For example, if the cations and anions are 
hydrophilic and $g_1\sim g_2\sim 15$, the shift term 
is of order $-500 n_0$ and its magnitude can be 
appreciable even for $v_0n_0\ll 1$.  
It is well-known that the coexistence curve 
and the spinodal curve of aqueous mixtures 
are much shifted even by a small amount of 
hydrophilic ions such as Na$^+$ and Br$^-$ \cite{polar}. 
(ii) On the other hand, $\gamma_{\rm p}$ can be increased 
for antagonistic salt \cite{OnukiJCP04,OnukiPRE,Sadakane}.  
From the last term in equation (3.16)  a Lifshitz point appears at   
$\gamma_{\rm p}=1$. (In figure 2, $\gamma_{\rm p}=
5/4\sqrt{3}<1$).  
That is, 
while $S(q)$ is maximum at $q=0$  for $\gamma_{\rm p}<1$, 
it  exhibits a peak  at an intermediate wave number 
 $q_{\rm m}=( \gamma_{\rm p}-1)^{1/2}\kappa$ 
  for $\gamma_{\rm p}>1$. 
The peak height is given by $S(q_{\rm m})=1/({\bar r}-r_{\rm m})$, where 
\be 
r_{\rm m} = (g_1+g_2)^2n_0/2 
- C(\gamma_{\rm p}-1)^2\kappa^2. 
\en 
For  ${\bar r}<r_{\rm m}$, 
mesophase formation should take place (see figure 4). 
For  ${\bar r}>r_{\rm m}$, two-phase states 
 with a planar interface can be  stable or metastable 
 even for $\gamma_{\rm p}>1$. 
(iii) If $\gamma_{\rm p}<1$, the criticality 
and the macroscopic phase separation can 
remain in the mean field 
theory. That is, for $q\ll \kappa$, 
we obtain the Ornstein-Zernike form, 
\be 
S(q)\cong  1/[{\bar r}- (g_1+g_2)^2n_0/2 + C(1-\gamma_{\rm p}^2)q^2].
\en 
The coefficient in front of $q^2$ is reduced 
by the factor $1-\gamma_{\rm p}^2$, 
which leads to enhancement of $S(q)$ 
in the range  $\xi^{-1}<q< \kappa$. 
 When  $\kappa>\xi^{-1}$, 
 the  correlation length $\xi$ 
should be redefined by  
\be 
\xi^{2}= C(1-\gamma_{\rm p}^2)/[{\bar r}- \frac{1}{2}(g_1+g_2)^2n_0]. 
\en

Second, retaining  the ion densities, 
we  eliminate 
 the composition  fluctuations in $F$ 
 by  setting 
\be 
({{\bar r}+C q^2}) \phi_{\small{\bi q}}= \sum_{i} 
g_i n_{i{\small{\bi q}}}, 
\en 
where the  terms nonlinear in $\delta\phi$ are neglected. 
We then obtain  effective  interactions 
among  the ions mediated by the composition fluctuations. 
The resultant free energy of ions is written as  
\bea 
&&{ F}_{\rm ion} = 
\int d{\bi r} 
  \sum_{i}  {T}n_i \ln (n_i\lambda_i^3)  \nonumber\\
&&\hspace{-1cm} +  \frac{1}{2} 
\int\hspace{-1mm} d{\bi r}\hspace{-1mm}\int\hspace{-1mm} d{\bi r}' 
 \sum_{i,j}  V_{ij}(|{\bi r}-{\bi r}'|) 
 \delta n_{i}({\bi r})\delta n_{j}({\bi r}' ) ,   
\ena 
where  the terms linear in $n_i$ are not written explicitly.  
Here the deviations 
$\delta n_i=n_i -\av{n_i}$ need not be very small and 
the logarithmic terms are not expanded with respect to them. 
The effective interaction 
potentials ${V_{ij}(r)}$ are expressed as \cite{OnukiPRE}
\be 
{V_{ij}(r)}= 
\frac{Z_iZ_je^2}{\ve r}-\frac{Tg_ig_j }{4\pi C}
\frac{1}{r}{e^{-r/\xi}} ,
\en 
where $Z_1e$ and $Z_2e$ are  the  ion charges 
(being $\pm e$ in the monovalent case) and  
$\xi= (C/{\bar r})^{1/2}$ is 
the correlation length without ions (different 
from  $\xi$ in equation (3.21)).

The solvation-induced interaction 
is effective in the range $a\ls r\ls \xi$ 
and can be increasingly 
important on approaching the solvent criticality (for $\xi\gg a$). 
(i) It is attractive among the ions of the same species $(i=j)$ 
and  dominates over the  Coulomb 
repulsion  for 
\be 
g_i^2>  4\pi C \ell_B  .
\en 
Under this condition   
there should be a  tendency of ion  aggregation 
of the same species.  (ii)In the antagonistic case 
($g_1g_2<0$), the cations and anions 
can repel one another in the range $a\ls r\ls \xi$  for
\be 
|g_1g_2|>  4\pi C \ell_B  ,
\en 
which triggers charge density waves near 
the solvent criticality.

The ion structure factors $G_{ij}(q)$ in equation (3.13) 
can be calculated from $F_{\rm ion}$ in equation (3.23). 
The inverse  of  the $2\times 2$ matrix 
$G_{ij}(q)$ is written as 
\be 
G^{ij}(q)= \delta_{ij}+ n_0 V_{ij}(q)/T, 
\en 
where $V_{ij}(q)$ is the Fourier transformation 
of $V_{ij}(r)$ in equation (3.24).  In the monovalent case, 
some calculations give desired  results, 
\bea 
&&\hspace{-3mm}\frac{G_{ii}(q)}{G_0(q)}=1 + {n_0}S(q)\bigg[ 
g_1^2+g_2^2- \frac{(g_1-g_2)^2}{2(u+1)}-
 \frac{2g_i^2u}{2u+1} \bigg],\nonumber\\
&&\hspace{-3mm}{G_{12}(q)}=\frac{1}{2} + \frac{n_0}{4}S(q) 
(g_1+g_2)^2- \frac{1}{4}C(q),  \nonumber\\
&&\hspace{-3mm}{C(q)}= \frac{2u}{u+1}+ {n_0}S(q)\frac{(g_1-g_2)^2}{(u+1)^2}u^2,
\ena  
where $u=q^2/\kappa^2$ and 
\be 
G_0(q)= \frac{u+1/2}{u+1} = 
\frac{q^2+\kappa^2/2}{q^2+\kappa^2}
\en 
is the structure factor for the cations (or 
for the anions) in the absence of 
solvation.  The solvation parts in equation (3.28)  
are all proportional to $n_0S(q)$, where $S(q)$ 
is given by equation (3.16), so they 
diverge on approaching the spinodal 
discussed below equation (3.18). The Coulomb 
interaction suppresses 
large-scale charge-density fluctuations, 
so $C(q)$ tends to zero as $q\to 0$.

\begin{figure}[t]
\begin{center}
\includegraphics[width=0.4\textwidth, bb= 0 0 574 450]{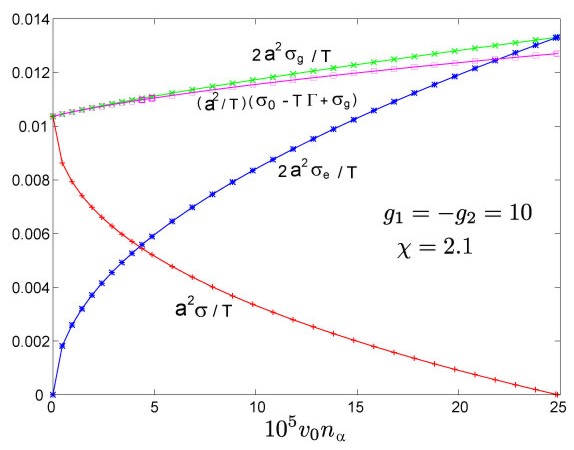}
\end{center}
\caption{Surface quantities $\sigma$, $2\sigma_e$, $2\sigma_g$, 
and $\sigma_0-T\Gamma+ \sigma_e$ in units of $Ta^{-2}$ 
vs $v_0 n_\alpha$ for a monovalent, antagonistic 
 salt with $g_1=-g_2=10$ at $\chi=2.1$. Here $2\sigma_g$ 
 and $\sigma_0-T\Gamma+ \sigma_e$  nearly coincide to 
 support equation (3.36).  
 }
\end{figure}

\subsection{Surface tension}
Addition of an antagonistic salt can  dramatically reduce the 
 surface tension $\sigma$ of 
 liquid-liquid interfaces even at small content  
\cite{OnukiJCP}. This trend was in fact observed  
\cite{Reid,Luo}.   
Here  the behavior of $\sigma$  
will be examined in some detail    
as a continuation of Ref.$[11]$, 
but the image 
interaction will be neglected.

We suppose a  planar liquid-liquid interface 
at $z=0$ taking the  $z$ axis in 
 its perpendicular direction.  
We impose the boundary condition  
$\phi \to \phi_\alpha$ in phase $\alpha$ 
and $\phi \to \phi_\beta$ 
in phase $\beta$ far from the interface and  minimize  
the grand potential $\Omega=\int d{\bi r}\omega$. 
The  grand potential 
density $\omega$ reads 
\be 
\omega=f- h\phi- \sum_i \mu_in_i, 
\en 
where $f$ is the free energy density  in equation (3.1) 
and the chemical potentials 
$h= \delta F/\delta \phi$ and 
$\mu_i=\delta F/\delta n_i$ are 
homogeneous constants. 
The relation   
$d\omega/dz= 2C\phi'\phi''- d(\rho \Phi)/dz$ is 
then   derived  and is  integrated to give  
\be 
\omega= C\phi'^2-\rho\Phi +\omega_\infty. 
\en 
where  $\phi'=d\phi/dz$ and $\phi''=d^2\phi/dz^2$.
Since $\phi'$ and $\rho$ tend to zero far from the interface, 
 $\omega(z)$ tends to 
a common value 
$\omega_\infty$ as $z\to \pm\infty$. 
The surface tension is written  as 
\be 
\sigma=\int dz [\omega(z)-\omega_\infty]
=  \int dz [ C\phi'^2- \frac{\ve }{4\pi}E^2].
\en

In our previous work \cite{OnukiPRE,OnukiJCP}, 
we expanded $\sigma$ 
with respect to the salt density as  
\be 
\sigma\cong \sigma_0 -T\Gamma - \int dz \frac{\ve }{8\pi}E^2,
\en 
where $\sigma_0$ is the surface tension without 
solute and 
$\Gamma$ is the surface adsorption 
of ions.  In terms of the total 
ion  density $n=n_1+n_2$, it is  
defined by 
\be 
\Gamma= \int dz[ n-n_\alpha -\frac{\Delta n}{\Delta\phi}
 (\phi-\phi_\alpha)],
\en 
where $n_K=n_{1K}+n_{2K}$ ($K=\alpha$ or $\beta$)  
and $\Delta n=n_\alpha-n_\beta$, so the integrand tends to zero 
as $z\to \pm\infty$. 
The expansion (equation (3.33)) is valid up to linear order 
in the salt density. 
 For   dilute neutral  solute, 
 it becomes the well-known Gibbs 
formula $\sigma\cong 
\sigma_0 -T\Gamma $ \cite{Gibbs}. The electrostatic contribution to 
 the surface tension has been neglected 
 in the Gibbs formula, but it can be crucial 
 for antagonistic salt \cite{OnukiJCP} 
 and for ionic surfactant \cite{OnukiEPL}.

We have numerically examined these relations. To this end, 
we  introduce the  areal densities of the electrostatic 
energy and the gradient free energy  as 
\be 
\sigma_{e}=   \int dz {\ve }E^2/{8\pi}, \quad 
\sigma_{g}= \int dz C\phi'^2/2.
\en 
Then equations (3.32) and (3.33)  yield 
$\sigma=2\sigma_g-2\sigma_e$  and  
\be 
2\sigma_g \cong \sigma_0 -T\Gamma +\sigma_{\rm e}. 
\en  
In figure 3, we show  the  surface quantities 
$2\sigma_e$,  $2\sigma_g$, 
$\sigma=2\sigma_g-2\sigma_e$, 
and the combination 
$\sigma_0 -T\Gamma +\sigma_{\rm e}$ as functions of 
the bulk ion density $n_\alpha$, 
where $g_1=- g_2=10$, $\chi=2.1$, and $\gamma_{\rm p}=3.455$.
Here  $n_\alpha=n_\beta$ for $g_1=-g_2$.
The surface tension vanishes 
at  $v_0n_\alpha= 2.5\times 10^{-4}$, where $\sigma_g=\sigma_e$. 
The  approximate relation  
(equation (3.36)) excellently  holds here, where 
 its  deviation is at most of order $10^{-3}T/a^2$.

\begin{figure}[t]
\begin{center}
\includegraphics[width=0.4\textwidth, bb= 0 0 533 525]{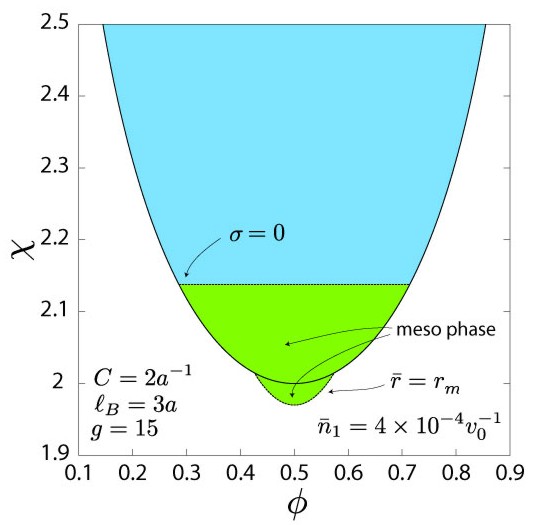}
\end{center}
\caption{Phase diagram for 
$g_1=-g_2=15$,  $v_0{\bar n}_1=4\times 10^{-4}$,  
and $\gamma_{\rm p}=1.727$ in the $\chi$-$\phi$ plane. 
Mesophases are realized in the  region (in green) 
where $\chi<2.14$ and ${\bar r}>r_m$. Macroscopic 
phase separation occurs 
in the two-phase region above the mesophase region.}
\end{figure}

\begin{figure}[t]
\begin{center}
\includegraphics[scale=0.36, bb= 0 0 503 350]{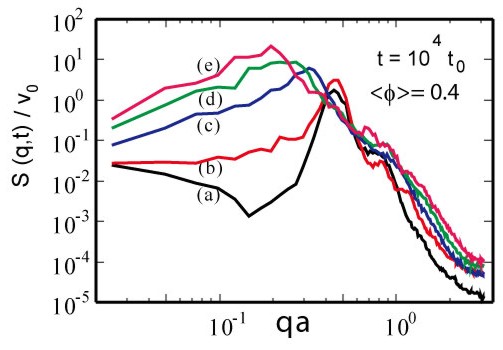}
\end{center}
\caption{
Structure factor $S(q,t)$ vs $q$ at $t=10^4t_0$
of  a binary  mixture  at $\av{\phi}=0.4$  
and $n_0v_0=0.002$ 
for (a)$\chi=2.0$, (b)2.1, (c)2.2, (d)2.3, and (e)2.4. 
}
\end{figure}


\begin{figure}[t]
\begin{center}
\includegraphics[scale=0.36, bb= 0 0 516 721]{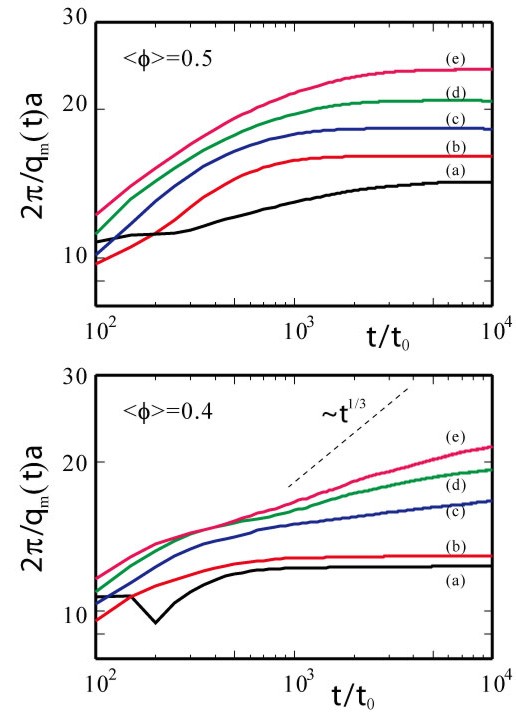}
\end{center}
\caption{
Characteristic domain size $2\pi/q_m(t)$ vs $t$  
of a binary  mixture  at $\av{\phi}=0.5$ (top) and 0.4 (bottom)  
for $n_0v_0=0.002$ 
for (a)$\chi=2.0$, (b)2.1, (c)2.2, (d)2.3, and (e)2.4. 
For $\av{\phi}=0.5$, the domain structure is nearly 
pinned at long times for these $\chi$. 
For $\av{\phi}=0.4$, the coarsening 
is nearly stopped for  $\chi=2.0$ and 2.1, 
while it is much slowed down for the larger $\chi$.  
As a guide, a line with slope $1/3$ is written (dotted line). }
\end{figure}

\begin{figure*}
\begin{center}
\includegraphics[width=0.8\textwidth, bb= 0 0 566 305]{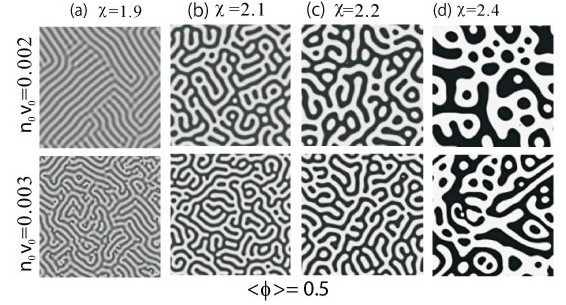}
\end{center}
\caption{
Composition  patterns 
of a binary mixture  at $\av{\phi}=0.5$  
containing an antagonistic salt 
 with  $v_0 n_0=0.02$ (top)  and 0.03 (bottom) at $t=10^4 t_0$, 
 where  (a)$\chi=1.9$, (b)2.1, (c)2.2, and (d) 2.4. 
 These patterns are frozen in time.  
}
\end{figure*}
\begin{figure*}
\begin{center}
\includegraphics[width=0.8\textwidth, bb= 0 0 591 333]{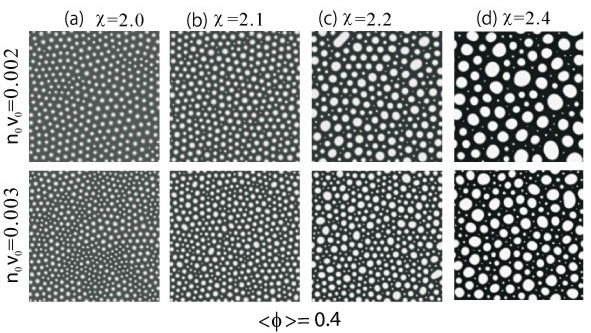}
\end{center}
\caption{
Composition  patterns 
of a binary mixture  at $\av{\phi}=0.4$  
containing an antagonistic salt 
 with  $v_0 n_0=0.02$ (top)  and 0.03 (bottom) at $t=10^4 t_0$, 
 where  (a)$\chi=2.0$, (b)2.1, (c)2.2, and (d) 2.3. 
Patterns are fronzen  for (a) and (b) 
and are  very slowly evolving for (c) and (d).  Size dispersity 
of droplets is conspicuous.}
\end{figure*}

\subsection{Phase diagram with antagonistic salt}

With addition of antagonistic salt,  
 the consolute criticality of binary mixtures 
 disappears due to emergence of mesophases.   
However, we have not yet fully understood  
the phase behavior even in the mean field theory. 
 So far, we have found 
(i) instability of one-phase states 
for $\gamma_{\rm p}>1$ in the range ${\bar r}<r_m$, 
as discussed around equation (3.19),  and 
(ii) vanishing of the surface tension 
$\sigma$ with increasing the salt density as  in figure 3 
or on approaching the solvent criticality. 
Thus, in figure 4, mesophases 
should be formed in the  area 
with $\chi<2.14$ and ${\bar r}<r_m$ 
in the $\phi$-$\chi$ plane. 
Here $g_1=-g_2=15$ at $v_0{\bar n}_1=
4\times 10^{-4}$.

\begin{figure}[t]
\begin{center}
\includegraphics[width=0.5\textwidth, bb= 0 0 540 728]{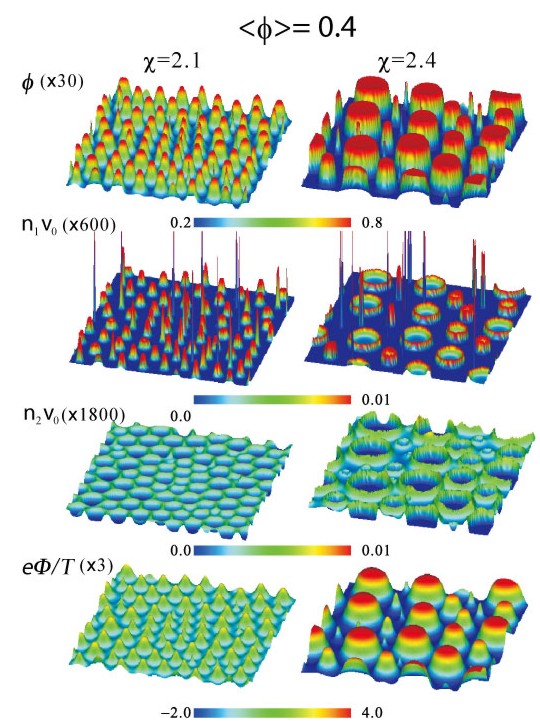}
\end{center}
\caption{Composition $\phi$, cation density 
$n_1$, anion density $n_2$,
 and  potential $\Phi$ (from top to bottom) 
for $\chi=2.1$ (left) and $\chi=2.4$ (right)   
in a binary  mixture  at $\av{\phi}=0.4$  
and $n_0v_0=0.002$. Snapshots  in a 
$1/4$ of the total system 
are displayed. 
}
\end{figure}

\begin{figure}
\begin{center}
\includegraphics[scale=0.36, bb= 0 0 527 744]{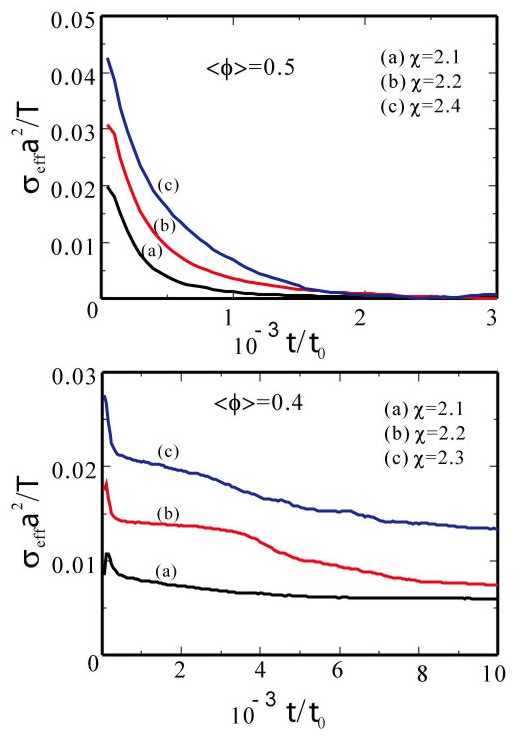}
\end{center}
\caption{
Effective surface tension 
 $\sigma_{\rm eff}$ 
 in equation (4.3) vs $t$ 
 for $\av{\phi}=0.5$ (top) and 0.4 (bottom). 
 It tends to zero  for $\av{\phi}=0.5$ 
 but  remains positive for $\av{\phi}=0.4$.
 }
\end{figure}

\section{Simulations in Two Dimensions 
with Antagonistic Salt}
\setcounter{equation}{0}

We already presented 2D simulation 
results of phase separation 
of a binary mixture with a monovalent,  antagonistic salt 
at the critical composition 
$\av{\phi}=0.5$ \cite{Araki}.   
Here we present further results 
in two dimensions  at  $\av{\phi}=0.5$ and 0.4. The composition 
 patterns are analogous to those 
 well-known for block copolymers, 
surfactant systems, and magnetic fluids 
\cite{Onukibook,Seul,Ohta}.  
In our ionic case, however, 
we shall encounter some unique features  
 in the droplet patterns. 
 We  stress  that three dimensional simulations 
 are needed to really understand the 
 phase ordering in this system. 
 For example, we should  reproduce 
 the observed onion  structure \cite{SadakanePRL}. 

\subsection{Background of simulation}

 We vary the interaction 
 parameter $\chi$  and set the average ion  density 
  $n_0= \av{n_1}=\av{n_2}$ equal to 
   0.002$v_0^{-1}$ or 0.003$v_0^{-1}$. 
The  other parameters are  
the same as in our previous 
work \cite{Araki}. 
Namely, we  set  $ C=a^{-1}$, $g_1=-g_2=15$, 
$\ve_1=0$, and $\ell_{\rm B}=3a$.  
Then we obtain  $\gamma_{\rm p}\cong 2.44$ from equation (3.18). 
These values of $g_1$ and $g_2$ are realistic 
in view of the data of the Gibbs transfer free energy,  
as discussed below equation (2.11). 
On a  $256\times 256$ square lattice, 
we  integrated  the dynamic equation (3.7) 
using the Bragg-Williams free energy density 
$f_0$  in equation (3.1), the Stokes approximation 
for the velocity field $\bi v$,  
and  the modified Poisson-Boltzmann 
expressions (equation (3.12)) 
for the ion densities $n_1$ and $n_2$.  
At $t=0$, we started  with  the initial condition 
$\phi({\bi r},0)=\av{\phi}$  with small random 
numbers superimposed. Subsequently, phase ordering took place for 
the parameter values given below. 
The space mesh size is  $a=v_0^{1/3}$ 
and  the time step of integration is 
$0.01t_0$, where $t_0$ is related to 
the kinetic coefficient $L_0$  in equation (3.7). 
and the viscosity $\eta$ in equation (3.9) as 
\be 
t_0= a^5/L_0, \quad L_0={0.16a^4 T}/\eta.
\en 
Time-evolution of the composition is 
 meassured in units of $t_0$. 
In late stages it is much slower than $t_0$.

\subsection{Slow coarsening 
and patterns}

In figure 5, we plot the  (angle-averaged) structure factor 
$S(q,t)$ of the composition 
vs $q$ for $\av{\phi}=0.4$ at $t=10^4t_0$.  It exhibits 
a sharp peak at $q\sim q_m(t)$. In our previous work 
\cite{Araki}, a similar plot of $S(q,t)$ was given 
for $\av{\phi}=0.5$. 
For each simulation run, we calculate 
the characteristic wave 
number  $q_m(t)$  by   
\be 
q_m(t)=\sum_{\small{\bi q}} |{\bi q}|
|{\bi \phi}_{\bi q}|^2
/\sum_{\small{\bi q}} |{\bi \phi}_{\bi q}|^2,  
\en  
where ${\bi \phi}_{\bi q}$ is the Fourier component of 
$\phi({\bi r},t)$. In figure 6, we plot the characteristic 
domain size $2\pi/q_m(t)$ vs $t$. 
For all $\chi$ at $\av{\phi}=0.5$ 
and for $\chi=$2.0 and 2.1 at $\av{\phi}=0.4$, 
the  coarsening is nearly stopped at long times. 
For $\chi\ge 2.2$ at $\av{\phi}=0.4$, 
it is much slowed down at long times and, if 
we write  $ q_m\sim t^\gamma$, the exponent $\gamma$   
increases from zero up to 
about $0.1$ at  $\chi =2.4$.   Note that the 
typical domain size grows as $t^{1/3}$ for 
conserved order parameter without the hydrodynamic 
interaction and is  faster 
with the hydrodynamic 
interaction\cite{Onukibook,Furukawa}.

We next show composition patterns at $t=10^4t_0$. 
From figure 6, they  are frozen or 
very slowly evolving.   
In figure 7,  they are  stripe-like and bicontinuous 
for  $\av{\phi}=0.5$.  In figure 8,   
 they are  droplet-like for $\av{\phi}=0.4$.  
The gradation of each snapshot 
represents the local value of $\phi({\bi r},t)$. 
For $\av{\phi}=0.5$ (or 0.4), 
the variance $[\av{(\phi-\av{\phi})^2}]^{1/2}$ 
is about 0.16 (or 0.13)  at $\chi=2$ and 
increases to 0.33 (or 0.31)  at $\chi=2.4$. 
Note that no phase separation occurs for $\chi \le 2$ 
without ions.  We recognize that 
 the patterns are very sensitive to the values of 
$\chi$ and $v_0 n_0$. 
A marked feature is that  the largest domain size  
increases with increasing $\chi$ 
and decreases with increasing $n_0$. More details are as 
follows.  (i) At  $\av{\phi}=0.5$ in figure 7,  the 
patterns  are lamellar-like 
for $\chi=1.9$, while they resemble those 
in systems with competing interactions 
\cite{Seul,Ohta}. 
(ii) In the droplet patterns at  $\av{\phi}=0.4$, 
water-rich  droplets  in a percolated  oil-rich  region  
 have   considerable size dispersity.  
This is possible only when even 
small droplets have long life times. 
(Without ions, small droplets quickly 
disappear due to the surface tension effect.)
Large and small droplets 
 segregate  exhibiting   
 polycrystal-like medium-range order for $\chi=2$, 
 while   larger droplets are  
growing and smaller ones are 
diminishing very slowly in random  
configurations for $\chi\gs 2.2$.

In figure 9, at  $\av{\phi}=0.4$,  we display 
the composition $\phi({\bi r},t)$, 
the ion densities $n_1({\bi r},t)$ and $n_2({\bi r},t)$, 
and the potential $\Phi({\bi r},t)$ 
at $t=10^4t_0$ for the two cases $\chi=2.1$ and 2.4.  
Similar snapshots at  $\av{\phi}=0.5$ 
can be found in our previous paper \cite{Araki}. 
The typical droplet radii are a few times larger 
for $\chi=2.4$ than for $\chi= 2.1$, 
but small droplets (with diameter$\sim 5a$) 
still remain in the two cases 
giving rise to sharp peaks 
in the cation distribution. Due to the strong 
selective  solvation, 
the cations are within the droplets and 
 the anions are outside them. 
For  $\chi= 2.1$, the typical  droplet size 
 is small and  the cation density 
is highest at the droplet centers.   
For  $\chi= 2.4$, the cations 
and the anions are mostly localized 
in the electric double layers at the  
interfaces.  In these cases, 
 the water-rich droplets are positively charged 
and are repelling  one another. 
However, we have not yet examined how this 
repulsion can influence the coarsening process.

 \subsection{Effective surface tension}
 
In figure 10, we show 
the time-development of 
the effective surface tension 
defined by 
\be 
\sigma_{\rm eff}= \frac{1}{S}\int d{\bi r}(C|\nabla\phi|^2- 
\frac{\ve}{4\pi} {\bi E}^2), 
\en 
where $S$ is the total interface length 
 calculated separately. 
This quantity becomes the usual surface tension for a planar 
interface in equilibrium from equation (3.32).  
(In our previous work \cite{Araki}, 
the space averages of the gradient free energy density
and the electrostatic energy  density 
were  separately plotted for $\av{\phi}=0.5$ 
in figure 10 there.)  
We recognize that 
$\sigma_{\rm eff}$ tends to zero for bicontinuous 
domain morphology but remains positive 
for droplet morphology. 
It is rather surprising that  the coarsening 
in the droplet case is much slowed 
down even for positive $\sigma_{\rm eff}$. 
To explain  these results theoretically,   
we need more systematic analysis of the problem.

\section{Summary and  remarks}

In this work, we  have demonstrated 
the crucial role of the preferential solvation 
of ions in phase transitions 
of polar binary mixtures. 
Together with a short summary of our previous work, 
we have presented some new results. 
(i) In equation (3.28), we have calculated the structure factors 
among the ion density fluctuations. 
(ii) In figure 3, we have examined how  the surface 
tension tends  to zero with increasing the  density 
of   antagonistic salt. 
(iii) In figure 4, we have shown a phase diagram 
of a binary mixtures containing antagonistic salt, 
where mesophases are realized near the  
solvent criticality. Note that the original criticality 
without ions disappears 
when antagonistic salt with $\gamma_{\rm p}>1$ is added. 
(vi) In figures 5-10, we have numerically examined 
the phase ordering of a binary mixture 
containing antagonistic salt 
for four values of $\chi$ 
in the bicontinuous case 
at  $\av{\phi}=0.5$ and in the droplet case at 
$\av{\phi}=0.4$. 
The coarsening is markedly slowed down  
with decreasing $\chi$ to 2 and/or 
with increasing the ion density $n_0$. 
The  size dispersity of 
water-rich  droplets is significant, 
because even small droplets 
have very long life times. 
In figure 9, we have found 
that water-rich  droplets are positively charged 
with the anions in the  percolated 
oil-rich region. In figure 10, the effective surface 
tension $\sigma_{\rm eff}$ vs $t$ is plotted for 
$\av{\phi}= 0.5$ and 0.4. It tends to zero 
for $\av{\phi}=0.5$ but 
remains positive for $\av{\phi}=0.4$.

In figures 7 and 8, 
the domain  configurations 
exhibit some mesoscopic order (lamellar-like or   
polycrystal-like)  for $\chi$ close to 2, while 
  they remain  disordered  for larger $\chi$ 
in our simulation time. In this paper, we cannot  conclude whether or not 
the domain growth has ultimately 
 stopped  or not.   If the 
thermal noise is present, 
the system might be slowly approaching  
lamellar or  crystalline  states  at very long times.

In our theory,   the molecular volumes 
of the two components are commonly $v_0$. 
However, the molecular volumes of 
 D$_{2}$O and 3MP (the inverse densities of 
the pure components) are 28 and 
 168 \AA$^3$, respectively, for example.
Moreover, the coefficient $C$ 
of the gradient free energy 
is  an arbitrary constant. 
Therefore, our theory remains very 
qualitative. 
It is not clear how $C$ can be 
determined in  aqueous mixtures 
where the hydrogen bonding is of primary importance.  
For pure water, 
the observed surface tension  outside the critical 
region  $1-T/T_c \gs 0.1$ 
 fairly agrees with the calculated  surface tension 
from the van der Waals  model  with 
the gradient free energy density 
$10 Ta^5 |\nabla n|^2/2$ \cite{Kiselev,Kitamura}, 
where $n$ is the density 
and $a=3.1{\rm \AA}$ 
is the van der Waals radius.

There can be a number of dynamical effects 
in the presence of antagonistic salt, 
including  multi-exponential 
decays of   the dynamic light scattering 
intensity \cite{Araki} and  
singular rheological behavior with 
appearance of mesophases \cite{Onukibook}. We  
propose experiments of the surface instability 
taking place for negative surface tension. 
That is, a water-oil interface 
becomes unstable with addition of  antagonistic salt 
above a critical amount (as suggested by figure 3) 
or on approaching the solvent criticality 
with a fixed density  of antagonistic salt. 
In the introduction, we have already 
mentioned the experiments 
of spontaneous emulsification 
at a water-oil interface 
with antagonistic salt \cite{Aoki,Poland}.

Though this paper has mostly treated antagonistic salt, 
intriguing solvation effects can well  be predicted 
for  aqueous mixtures containing 
 usual hydrophilic ion pairs. 
For example,  many  groups have detected  long-lived 
heterogeneities by dynamic light scattering 
in one-phase states of various 
aqueous mixtures containing hydrophilic salt \cite{So}. 
Recently, we have shown that 
preferential solvation 
can stabilize water-rich   domains enriched with ions even for 
$\chi<2$ when  water is the  minority component 
\cite{precipitation}. We also mention 
the problem of preferential wetting  of  mixture solvents 
on ionizable surfaces,  which  can be either  planar, spherical, 
or cylindrical. Around 
 an ionizable  rod, we  already predicted  a first-order  
wetting  transition   with 
discontinuous jumps   in the preferential adsorption and 
the degree of ionization \cite{Oka}.  
In such situations, 
we should take into account 
the  solvation interaction among the charges (consisting of the ionized 
units on the surface  and the counterions) 
and the polar molecules.

Finally, it is worth noting that 
ion-induced nucleation 
has long been investigated 
in the literature \cite{Thomson,Kusaka}.
It is well-known that 
water droplets  can easily be 
produced  around hydrophilic ions 
in metastable gas mixtures containing water vapor.
Here ions play the role of 
nucleation seeds on which hydration-induced 
 condensation occurs. 
A Ginzburg-Landau approach to this problem 
in the line of Ref.$[9]$ 
was also presented \cite{Kitamura}.

\vspace{2mm}
\noindent{\bf Acknowledgments}\\
\noindent 
This work was supported by KAKENHI (Grant-in-Aid for Scientific Research)
on Priority Area  Soft Matter Physics from
the Ministry of Education, Culture, 
Sports, Science and Technology of Japan.
Thanks are also due to K. Sadakane and H. Seto for 
informative discussions.

\end{document}